\def\year{2019}\relax

\documentclass[letterpaper]{article} 
\usepackage{aaai19}  
\usepackage{times}  
\usepackage{helvet} 
\usepackage{courier}  
\usepackage[hyphens]{url}  
\usepackage{graphicx} 
\usepackage{graphicx}  
\usepackage{CJKutf8}

\title{GenerationMania: Learning to Semantically Choreograph}

\author{
Zhiyu Lin\footnote{Contact Author},
Kyle Xiao\and
Mark Riedl\\
Georgia Institute of Technology\\
\{zhiyulin; kylepxiao\}@gatech.edu; riedl@cc.gatech.edu\\
}

\begin{document}
\maketitle

\begin{abstract}
Beatmania is a rhythm action game where players must reproduce some of the sounds of a song by pressing specific controller buttons at the correct time.
In this paper we investigate the use of deep neural networks to automatically create game stages---called charts---for arbitrary pieces of music. 
Our technique uses a multi-layer feed-forward network trained on sound sequence summary statistics to predict which sounds in the music are to be played by the player and which will play automatically. 
We use another neural network along with rules to determine which controls should be mapped to which sounds.
We evaluated our system on the ability to reconstruct charts in a held-out test set, achieving an $F_1$-score that significantly beats LSTM baselines.
\end{abstract}

\section{Introduction}

Rhythm action games such as {\em Dance Dance Revolution}, {\em Guitar Hero}, or {\em Beatmania} challenge players to press keys or make dance moves in response to audio playback.
The set of actions the player is required to do timed to the music is referred to as a {\em chart} and is presented to the player as the music plays.
Some rhythm action games involve the reconstruction of music where some sound events are played when the player performs the right action at the right time and other sound events are played automatically. 
This includes {\em Guitar Hero}---the player plays some of the guitar notes---and {\em Beatmania}---the player acts as a DJ that must play sound events from many different instruments; 
the music sounds incomplete or garbled if the player skips actions or misses the timing of actions.
Other games such as {\em Dance Dance Revolution} involve dance moves that are choreographed to the music but does not require a strict mapping between actions and sound events;
the music sounds complete regardless of what the player does.

The charts in rhythm action games are typically hand-crafted, limiting gameplay to songs that already have accompanying charts.
Because of this limitation, active novice communities have arisen to create new music and new accompanying charts. 
However, chart creation is considered a difficult and time consuming task. 
In this paper, we explore the use of deep neural networks to automatically construct charts for arbitrary pieces of music.
We target the {\em Beatmania IIDX} game because it has the oldest and most mature homebrew community, providing open-source game emulators and a larger corpus of charts and music than for other games. 

\begin{figure*}[tb]
\centering
\includegraphics[width=0.8\textwidth]{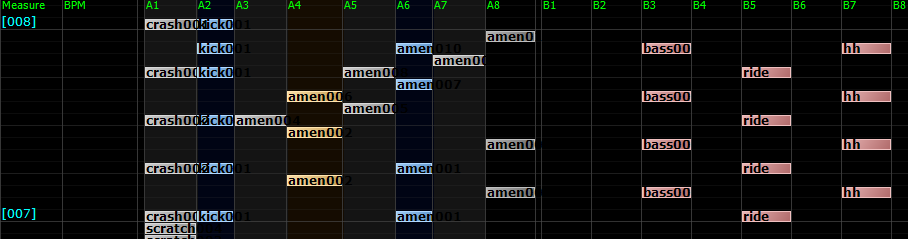}
\caption{A visualization of a single measure of the Beatmania IIDX homebrew chart for \textit{Poppin' Shower}. The `A' columns are for playable sound events for different controls and `B' columns are for non-playable sound events. The sound events are labeled with the names of the audio sample sound files.
Time progresses from bottom to top, corresponding to the way sound events ``fall'' in the game interface.}
\label{figure-gameplay}
\end{figure*}

Beatmania IIDX is a rhythm action game in which the player reconstructs the music by playing different sound events from different instruments at precise times.
Some sounds from some instruments are played automatically while others must be played by the player. 
That is, there are ``playable'' and ``non-playable'' sound events for each piece of music.
Playable sound events appear visibly in the chart to cue the player which action to invoke and have a one-to-one correspondence with an audio sample that will be heard in the music.
Non-playable sound events are considered part of the chart but are not shown to the player and the audio samples play in the background.
See Figure\ref{figure-gameplay} for an example chart.
In rhythm action game terminology, a {\em keysound} refers to the one-to-one mapping of a segment of music to a playable (or non-playable) chart sound event.\footnote{Beatmania IIDX is \textit{keysounded} by this definition; Dance Dance Revolution is \textit{non-keysounded} respectively.}

Our technique learns a model of how chart elements relate to underlying music.  Training the model consists of three tasks: (1)~analyzing the music sample to identify the instruments used; (2)~analyzing sample charts to identify relationships between sound events and the difficulty of charts per a variety of temporal scales; and (3)~learning a model to classify whether sound events are playable or non-playable.
To choreograph a new chart, the model is given a new music sample and produces a specification of which sound events should be playable and which should be non-playable.
A second neural network in conjunction with heuristic rules are then used to assign controls to each playable sound event.

Our work differs from other work on neural rhythm action game chart generation in three significant ways. 
First, other neural generative systems (e.g., {\em Dance Dance Convolution}~\cite{donahue2017dance}) target games that do not use keysounds---actions only need to follow the tempo without necessarily expressing any relationship to the underlying music. 
Second, other neural generative systems use recurrent neural networks, whereas we empricially find that a specialized multilayer feed-forward network that operates on a fixed window of chart elements at different timescales works best. 
Third, to support novice chart creators, we explicitly designed our chart generation technique to allow users to provide target difficulty progression and bias the generator's sytlistic choices.

We conduct an evaluation to measure the precision and recall of our technique in recreating the charts in a held-out test set.
Our system outperforms a number of strong baselines with respect to $F_1$-score, a harmonic mean of precision and recall, indicating ability to both determine when playable events should be in the chart and when non-playable events should be in the background.

In addition, we introduce the {\em BOF2011} dataset for Beatmania IIDX chart generation.\footnote{Our implementation and dataset is available at https://github.com/xxbidiao/GenerationMania}

\section{Background and Related Work}


%


The homebrew community of BMIIDX is arguably one of the oldest and most mature groups of its kind~\cite{chan2004cpr}, with multiple emulators, an open format (Be-music Source\footnote{\url{ http://www.charatsoft.com/develop/otogema/page/04bms/bms.htm}}, BMS) and peer-reviewed charts published in semi-yearly proceedings\footnote{\url{http://www.bmsoffighters.net/}}. 
Despite the community striving to provide the highest quality charts, due to the strict keysound requirements, usually the author of the music or a veteran chart author has to assist in the authoring of the chart.
Many aspiring amateur content creators start by building charts for non-keysound rhythm action games (i.e. {\em Dance Dance Revolution} chart building is considered by the community to be easier to learn). 
Furthermore, there is a strong demand for customized charts: players have different skill levels and different expectations on the music-chart translation, and such charts are not always available to them.





There are a handful of research efforts in chart choreography for rhythm action games, including rule-based generation~\cite{o2003dancing} and genetic algorithms using hand-crafted fitness functions~\cite{nogaj2005genetic}.
Procedural Content Generation via Machine Learning (PCGML) \cite{summerville2018procedural} is an emerging field of study.
Under its umbrella, {\em Dance Dance Convolution}~\cite{donahue2017dance} is the first deep neural network based approach to generate rhythm action game charts. 
%
Dance Dance Convolution uses a two-stage approach.
{\em Onset detection} is a signal analysis process that determines the salient points in an audio sample (drum beats, melody notes, etc.) where actions should be inserted into a chart.
{\em Action selection} uses a long-short term memory neural network~\cite{hochreiter1997long} to learn to map onsets to specific chart elements.
However, while onset detection was useful for action selection, other features such as music pitch did not significantly improve accuracy.
Our work on Beatmania IIDX chart generation differs from Dance Dance Convolution in that the primary challenge is determining whether each sound event should be playable or non-playable.
A recent work, \cite{fukunaga}, also utilized machine learning to classify playable events.
They used audio and timing features of current and 10 previous events, while we summarized these events along with other features and put them into an end-to-end pipeline.


%


\section{Dataset}

We compiled a dataset of songs and charts from the ``BMS Of Fighters 2011'' community-driven chart creation initiative.
During this initiative, authors created original music and charts from scratch. 
The dataset thus contains a wide variety of music and charts and was composed by various groups of people. 
Most authors build 3 to 4 charts for each song. 
The dataset, which we refer to as ``BOF2011'', consists of 1,454 charts for 366 songs. Out of 4.3M total sound events, 28.7\%, or 1.24M of them, are playable ones.
Table~\ref{table-dataset} summarizes the dataset.

\begin{table}
\caption{BOF2011 dataset summary.}
\label{table-dataset}
\centering
\begin{tabular}{l|l}
\hline
\# Songs                & 366\\
\# Charts               & 1,454\\
\# Charts per song      & 3.97\\
\# Unique audio samples & 171,808\\
\# Playable objects     & 1,242,394\\
\# Total objects        & 4,320,683\\
Playable object \%          & 28.7\\
\hline
\end{tabular}
\end{table}

We find that modeling the difficulty of charts plays an important role in learning to choreograph new charts; this observation is also made by~\cite{donahue2017dance}.
However, We discarded human-made difficulty labels due to its ambiguity.

Most of the charts in our dataset are relatively easy, in which non-playable sound events dominate. 
Furthermore, many of the same samples are repeatedly used, such as drum samples placed at nearly every full beat throughout a chart, resulting in only 171k unique audio samples in our dataset. 

\begin{figure*}
\includegraphics[width=\textwidth]{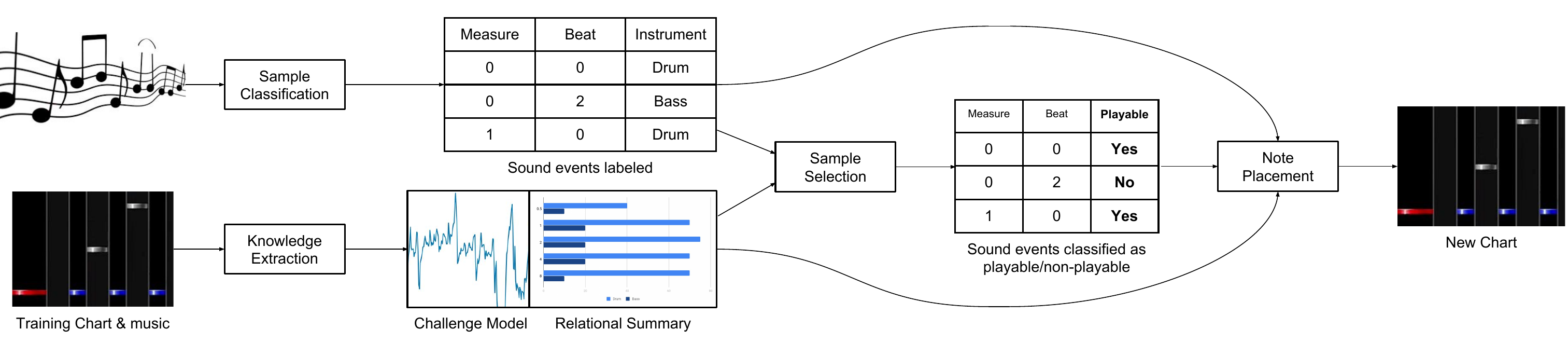}
\caption{The chart generation pipeline.}
\label{figure-pipeline}
\end{figure*}



\section{Methods}

Our chart generation system for BeatMania IIDX learns a model that relates audio samples to chart elements.
There are three tasks that must be accomplished to learn this model:
\begin{enumerate}
\item{\textit{Sample Classification}---Identifying the instrument used in audio samples.}
\item{\textit{Chart Knowledge Extraction}---Establish structure of each part in a chart including how sound events relate to each other and difficulty over time.}
\item{\textit{Sample Selection}---Classifying sound events as playable and non-playable.}
\end{enumerate}
Tasks 1 and 2 reduce each music sample and accompanying chart to a set of sound events and summary information containing event timing, difficulty, and the pattern of instrument probabilities across different time scales (see Section~\ref{sec:keg}).
Task 3 uses the summary information to train a model that categorizes sound events as playable or non-playable.

Chart generation involves transforming a new music in the same way and predicting the playability of each sound event. 
Once each sound event has been categorized as playable or non-playable, a fourth task, {\em Note Placement}, assigns each playable sound event to a control. 
See Figure~\ref{figure-pipeline}.


\subsection{Sample Classification}
\label{section-sample-classification}
Sample classification is a process by which notes from different instruments in the audio samples are identified.
The BMS file format associates audio samples with timing information.
That is, a chart is a set of sample-time pairs containing pointers to the file system where an audio file for the sample resides.
Unfortunately, in the BMS file format there is no standard for how audio samples are organized or labeled. 
However, many authors do name their audio sample files according to common instrument names (e.g., ``drums123.ogg''). 
The goal of sample classification is to label each sample according to the instrument based on its waveform. The predicted labels will be used to create one-hot encodings for each sample for the sample selection stage on the pipeline.

We construct a training set by gathering audio samples together with similar instrument names according to a dictionary and use the most general instrument name as the supervision label. We use the 27 most common categories for labeling, which was determined to work well experimentally.
To ensure that we don't overfit our classifier we train on an alternate dataset, ``BMS of Fighters Ultimate'' (BOFU) that does not share any music or charts with BOF2011, with a partially labeled dataset containing 60,714 labeled samples. Not every audio sample has a classifiable name, which we count as unlabeled.

\begin{figure*}
\includegraphics[width=\textwidth]{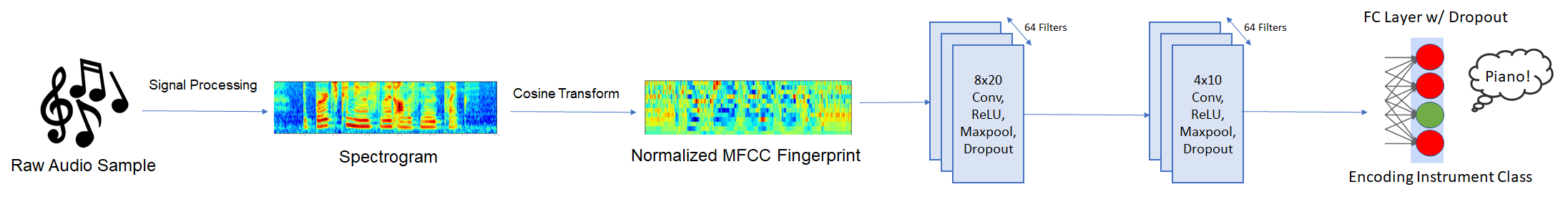}
\caption{The Sample Classification Pipeline.}
\label{figure-example}

\end{figure*}

We process the audio samples into a spectrogram representation and transform it into ``audio fingerprints,'' which is a vector of mel-frequency cepstral coefficients (MFCC) \cite{logan2000mel} of the log-magnitude mel-scale frequencies over time. This mapping is done via a linear cosine transform. We also fix the bit rate of the sound to 16k, so that the representation has a consistent temporal resolution. This pipeline is shown in Figure~\ref{figure-example}.

For our model, we followed the method described in ~\cite{sainath2015convolutional}. We feed the fingerprints through two 2D convolutional layers with bias. Each of the layers is followed by a Rectified Linear Unit (ReLU) activation function followed by max-pooling. Finally, we feed the results into a fully connected layer, which then outputs a one-hot encoding of the predicted category. We use a gradient descent optimizer with 50\% dropout rate and a step size of 0.01. 

After training on the BOFU dataset, we achieved an 84\% accuracy based on a 80-10-10\% train-test-validation split.

\subsection{Chart Knowledge Extraction}
\label{sec:keg}

Inspired by community chart authoring practices and music theory we provide an abstracted blueprint of charts instead of raw chart file formats.
This blueprint technically abstracts the chart from the underlying music, but references the instrument classes of each sound event, at different music timescales.
This blueprint consists of three types of knowledge that is automatically extracted from a chart: \textit{beat phase} capturing timing of sound events, \textit{challenge model} capturing the relative difficulty sections of the chart, and a \textit{relational summary} that captures the relationship between sound events in the chart.

\textit{Beat phase} denotes the timing relationship between a sound event and its underlying music as a whole.
We separate each interval representing a 4th musical note into sixteen typically 20ms windows.
We assign a value between 0 and 15 for each window, 0 denoting the note being on a down beat, 8 for halfway between two down beats (8th), and so on.
This yields a value for each sound event.

We find the relative difficulty of the portion of a chart that a sound event occurs in to be an important feature in determining whether a sound event should be playable or not. 
Our \textit{challenge model} describes how difficult each part of a chart will be perceived by a human.
We use a rule-based technique that has been used for other rhythm action games.
\footnote{Adapted from the Osu! framework. Their reference implementation is available at  \url{https://github.com/ppy/osu/tree/master/osu.Game.Rulesets.Mania/Difficulty}}
We first calculate a {\em strain} value for each playable event, then max pool it over 400ms intervals.
The strain value of each sound event is calculated as the
weighted sum of {\em individual strain} and {\em overall strain}.
Individual strain is the interval between sound events mapped to the same control on an exponentially decaying scale such that short intervals have higher strain values than long intervals. 
Overall strain is based on the number of controls that must be activated simultaneously.
We apply the pooled strain values back to each sound event, and get a value for each event.

\begin{figure}[h]
    \centering
    \includegraphics[width=2.5in]{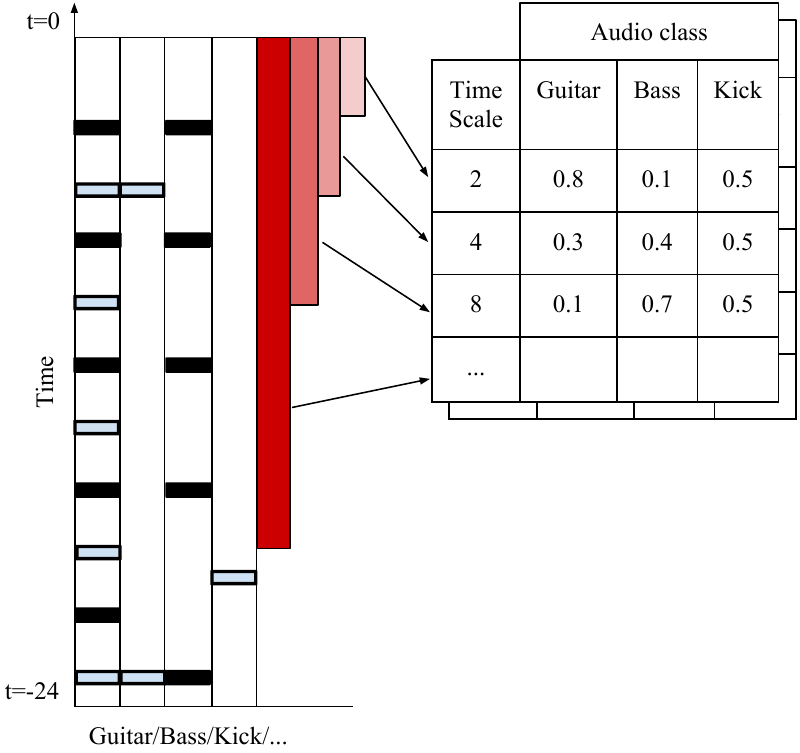}
    \caption{Depiction of a relational summary for a sound event at time $t=0$. Black bars denote playable sound events and blue bars denote non-playable ones in the chart. Red bars indicate windows of chart history at different timescales.}
    \label{fig:summary}
\end{figure}

The \textit{relational summary} draws a big picture of the chart in different scales.
Summarization is a technique popularized by {\em WaveNet}~\cite{van2016wavenet} to factor prior information from a fixed window at different time scales into a current prediction.
Unlike {\em WaveNet} which operate with inputs at the most fine-grained level of detail, we \textbf{summarize} previous events in feature space.
We found out that time scales of size 2, 4, 8, 16, and 32 beats best suit our purpose of summarizing both local and broad features.
For each discrete point in time in each timescale we compute the probability that a sound event resides in this scale of each instrument class which is described in section \ref{section-sample-classification}.
We illustrate relational summaries in Figure~\ref{fig:summary}.
Considering a sound event at time $t=0$, we compute the likelihood that that sound event will be playable or non-playable when considering different possible sample classes and looking back over a window of time of varying time scales. 
Thus, each sound event is represented by an $S \times C \times 2$ matrix where $S$ is the number of time scales and $C$ is the number of sample classes.

For each chart the beat phase information, challenge model, and relational summary forms a {\em summary matrix}.

l
\subsection{Sample Selection}
\label{sec:sample-selection}

Sample selection is the task of determining which sound events in the music should be playable and which should be non-playable.
Our Sample Selection prediction model is a feed-forward network consisting of 4 fully connected layers of dimensions 64, 32, 16, and 2, each followed by a Rectified Linear Unit (ReLU).
To perform sample selection, we pick the output node with the highest activation corresponding to the playable or non-playable classes.
Due to a class imbalance because most of the sound events are non-playable, we found that a weighted Mean Squared Error loss function helps improve the performance of the training.

At training time the summary matrix is derived from the training data so that the model can be trained to reconstruct the input data. 
At generation time, a summary matrix must be provided.
When generating charts for new music, the challenge model and relational summary can be taken from other charts.
The relational summary acts similar to a chart authoring {\em style}, providing an indication of preference for certain types of sound events to be made playable.
For example, the relational summary from a chart with a lot of playable guitar events will bias our system toward the same.
The chart from which to take challenge models and relational summaries do not necessary need to be in the dataset.
It is even possible to hand-author challenge models and relational summaries to induce a specific difficulty profile and style.





\subsection{Note Placement}

Note placement occurs during chart generation, once we have classified each sound event as playable or non-playable.
Each {\em playable} sound event at each timestep must be mapped to one of 8 controls. 

Any process that doesn't map objects to the same control at the same time is sufficient to make a chart playable, thus note placement is not a significant contribution in this paper.
We use a feed forward neural network similar to that in Section~\ref{sec:sample-selection} except the final layer is a softmax prediction over the 8 controls.
We post-process the note-placement with a heuristic, developed based on common chart design patterns.
If more than two sound events are mapped to the same control at nearly the same time, we instead place it in mirrored position (i.e. Left-most become right-most, etc.)
If that position is still occupied, we shift it to right-hand-side adjacent control. 
We repeat the adjacent control shift until an available position is found. 
Very rarely, if all 8 position is occupied, we label this sound event as unplayable and discard this event from the note placement task.


\begin{table*}[tb]
\centering
\footnotesize
\caption{Results for playable classification experiments, presented in mean and standard deviation. BP, CM and RS refer to Beat Phase, Challenge Model, and Relational Summary, respectively. $N=145$ for the test set.}
\label{table2}
\begin{tabular}{l|c|c|c}
\multicolumn{1}{c}{\textbf{Model}}                 & \multicolumn{1}{c}{\textbf{$F_1$-score}} & \multicolumn{1}{c}{\textbf{Precision}} & \multicolumn{1}{c}{\textbf{Recall}} \\
\hline
\multicolumn{4}{l}{\textbf{Reference Baselines}}                                                                       \\\hline
Random                                             &$0.291\pm0.089$&$0.335\pm0.200$&$0.299\pm0.020$\\
All Playable                                       &$0.472\pm0.207$&$0.335\pm0.199$&$1.000\pm0.000$\\
\hline
\multicolumn{4}{l}{\textbf{LSTM Baselines Models}}                                                          \\\hline
LSTM + Audio Features + BP + CM           &$0.424\pm0.154$&$0.767\pm0.176$&$0.353\pm0.248$\\
LSTM + Audio Features + BP + CM + RS &$0.564\pm0.149$&$\mathbf{0.776\pm0.117}$&$0.475\pm0.194$\\
\hline
\multicolumn{4}{l}{\textbf{Feed-Forward Models}}                                                         \\\hline
FF + Audio Features + BP + CM&$0.253\pm0.143$&$0.523\pm0.266$&$0.179\pm0.113$\\
FF + Audio Features + BP + CM (Self Summary)   &$0.368\pm0.198$&$0.422\pm0.213$&$0.392\pm0.258$\\
FF + Audio Features + BP + RS         &$0.621\pm0.206$&$0.760\pm0.110$&$0.568\pm0.254$\\
FF + Audio Features + BP + CM + RS        &$\mathbf{0.700\pm0.158}$&$0.762\pm0.114$&$\mathbf{0.662\pm0.193}$\\
\hline
\end{tabular}
\end{table*}

\section{Experiments}

We evaluate variations of our sample selection model against a number of baselines.
We focus on playable/non-playable classification because any reasonable note placement will work from the player's perspective.
We used a supervised evaluation metric: we embed a Summary Matrix extracted from the ground truth chart and then measure how accurate sample selection is compared to the original chart.
This is a necessarily artificial test in the sense that we have a ground-truth when comparing charts back to those in the testing set.

We establish two guidelines for a good generation model: it should not only predict \textit{playables} when they should be presented to players (high recall), but also ensure that playable events presented to players are actually playable events (high precision). 
A model achieving higher performance level on these two metrics demonstrates better ability to pick up essential sound events representing understanding of music resembling human author choices.
We applied an 80\%, 10\%, 10\% split on training, validation and testing data. 
Since the charts for the same music share similar traits, we ensured that such charts are not both in the training split and the testing split. 
We trained all models using the training split and report all results on the testing split.


We experiment with the following models:
\begin{itemize}
\item{{\bf Random Baseline:} classifies a given object as a playable with a probability of 0.3, chosen to give the best result;}
\item{{\bf All-Playable Baseline:} classifies all objects as a playable.}
\item{{\bf LSTM Baseline:} a seq2seq model \cite{NIPS2014_5346} with 
hidden and output layer size of 2. The highest activated output is selected as the prediction.}
\item{{\bf Full Model:} The sample selection model with summary matrix.}
\item{{\bf Self-Summary Model:} the same as the full model except that an empty summary matrix is initially provided and generated row-by-row on the fly based on previous generated predictions.}

\end{itemize}

\noindent 
The LSTM baseline was chosen because the conventional wisdom, established by Donahue et al.~\shortcite{donahue2017dance}, is that an LSTM is needed to model chart progression.
However, it is impossible to directly compare our model to the approach used in Donahue et al. because of the difference between keysound-based and non-keysound-based rhythm action games.

In our experiments, we explore different combinations of input features drawn from the summary matrix.
We developed our Neural Network models in PyTorch.
We used a mini-batch of 128 for the feed forward model; Due to the need for processing very long sequences with high temporal resolution, the LSTM model is trained by each sequence and is run in CPU mode.
The Full model converges in around 2 hours in GPU mode while the LSTM model takes far longer at around 50 hours, on a single machine using Intel i7-5820K CPU and NVIDIA GeForce 1080 GPU.

\section{Results and Discussion}


The experiment results are shown in Table~\ref{table2}. We only show configurations that beat reference baselines.
Our primary measure is the $F_1$-score, a harmonic mean of precision and recall. 

The first observation is that, in all cases, relational summary features boost performance across all models that use them.
The Full Model with relational summaries performs better than an LSTM model without relational summaries.
This is notable because an LSTM makes use of history and attempts to model the relationship between prior time steps and the next time step. 
The Full Model does not make use of history directly but has indirect access in the form of the summary statistics.
We hypothesize the high variance in charts hampers LSTM making it performs worse on our task.

Adding relational summaries to the LSTM model gives an LSTM access to history and also a hierarchy of statistics at different timescales.
The LSTM with relational summaries achieves a precision of $0.776\pm 0.117$ but only a moderate improvement in recall.
This is statistically not significantly better than the precision of the Full Model at $p > 0.025$ ($N=145$).
While this version of the LSTM arguably produced the highest precision it is at the expense of recall, meaning in practice that it produced a lot of charts where sound events are not predicted as playable where they should be
(perfect recall can be achieved by classifying all sound events as playable).
For a rhythm action game chart, one needs both high precision and high recall.

The Full Model with relational summaries achieves the highest $F_1$-score at $0.700 \pm 0.158$, which is our primary measure.
The Full Model $F_1$-score is statistically higher than that of the best LSTM model at  $p<0.00001$. 

\begin{figure}[t]
\includegraphics[width=0.5\textwidth]{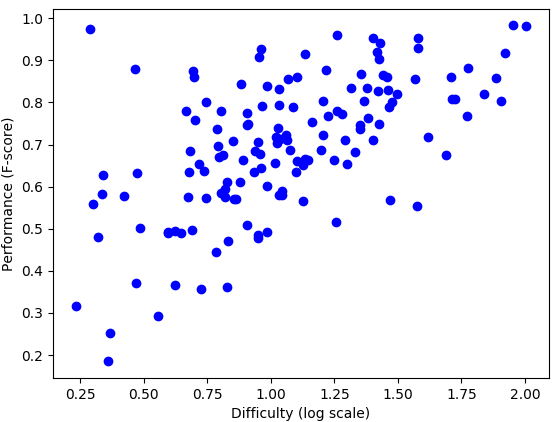}
\caption{The performance of feed forward model (with summary) regarding relative difficulty of the ground truth chart.}
\label{figure-diff-f1-feedforward}
\end{figure}

In our Full Model with relational summaries, the challenge model provides a $7.9\%$ improvement in the $F_1$-score.  
As with Donaue et al.~\shortcite{donahue2017dance}, we also observe that all models varied in performance on charts with different difficulty. %
We analyzed the effect of chart difficulty on our best performing model---the Full Model with relational summaries.
Plotting overall chart difficulty against $F_1$-score performance on each chart (Figure~\ref{figure-diff-f1-feedforward}) reveals that the Full Model achieves consistently high $F_1$-scores when the chart difficulty is high but has high variability in performance when chart difficulty is low.
There are fewer playable sound events in low-difficulty charts and we hypothesize that there is more variance to how authors select playable events.

The Full Model is naturally sensitive to the features. 
We also experimented with other features including sound event density, audio pitch, different numbers of instrument classes, and a hierarchical representation with musical measure as intermediate unit of organization.
These features failed to improve performance and in some cases reduced performance.
We also considered (a) an auto-encoder structure for LSTM model, which tries to auto-summarize the chart, (b) multi-layer LSTM structures more similar to that in Donahue et al.~\shortcite{donahue2017dance}, and (c) using different minimal unit of resolution, such as musical beats, in classification task. 
However, these models either overfit quite quickly or have unrealistic computational requirements.

As a side effect of how beat-phase information is organized in our specific task we are unable to include $\Delta$-beat features.
$\Delta$-beat measures the number of beats between the previous and next step and was used in Donahue et al.~\shortcite{donahue2017dance}.
The limitation stems from (1)~several semantically unrelated notes can be placed at the exact same time and (2)~notes can be placed in very short intervals (such as when representing a glissando). These issues prevent effective $\Delta$-beat detection in granularity of single notes. 

Unlike \textit{Sample Selection}, it is non-trivial to develop a quantitative method to evaluate Note Placement.
A similar metric which is used in evaluation of \textit{Sample Selection} would make little sense since no baseline Placement information is provided for non-playable notes. 
Furthermore, since this metric is per-note based, it will not award reconstruction of \textit{patterns} where groups of notes are placed in a certain shape, if the pattern is misplaced.
Due to these reasons, such a metric will offset drastically from human players' perception, rendering it ineffective.
This urges development of metrics for qualitative studies for Step Placement model.
Aside from that, our Challenge Model technique is completely heuristic based, meaning that it is sensitive to parameter tuning.
A heuristic-free approach or a model learned from player experience may help in this scenario.
We leave them as future work.


%
\section{Conclusions}

Choreographing rhythm action game charts is a challenging task. 
Beatmania is a keysounded rhythm action game, requiring chart elements to reflect the underlying music more closely than other, non-keysounded games.
We have established a technique for generating charts for new music samples that allows for user control of difficulty progression and also allows the user to bias the generator's stylistic decisions.
We also provide a new chart dataset for reproducible evaluations and to facilitate further research on rhythm action game choreography.

We show that a feed-forward network with a challenge model and relational summary performs better than strong LSTM baselines at determining when playable events should be in the chart and when non-playable events should be in the background.
Such a supervised evaluation is necessary to give confidence that sample selection is meaningful.
The full use case is for users to provide their own challenge models and relational summary information when creating charts for new music, or to create {\em blends} by borrowing this information from other, existing charts and applying it to new music.
Aside from solving a challenging creative task, intelligent systems that allow for user control---such as the technique described in this paper---can be of benefit to homebrew chart choreography communities by helping novices overcoming skill limitations.


\bibliographystyle{aaai}
\begin{CJK}{UTF8}{min}
\bibliography{refs}
\end{CJK}
\end{document}